\documentstyle[epsfig,longtable]{aipproc}

\def\la{\mathrel{\hbox{\rlap{\hbox{\lower5pt\hbox{${^\sim}$}}}\hbox{\lower0.5pt\hbox{${^<}$}}}}}
\def\ga{\mathrel{\hbox{\rlap{\hbox{\lower5pt\hbox{${^\sim}$}}}\hbox{\lower0.5pt\hbox{${^>}$}}}}}
\def\kmsmpc{km s$^{-1}$ Mpc$^{-1}$}
\def\lg{\mathop{{\rm log}_{10}}}

\begin{document}

\title{Evolution of the 1.4~GHz Radio Luminosity Function}
 
\author{Ian Waddington} 
\address{Department of Physics \& Astronomy, Arizona State University,
Tempe AZ 85287--1504, USA, and The Institute for Astronomy, University
of Edinburgh, Royal Observatory, Blackford Hill, Edinburgh EH9 3HJ,
UK.}

\maketitle

\begin{abstract}
The results of an optical and infrared investigation of a complete
sub-sample of the Leiden-Berkeley Deep Survey are presented.  Optical
counterparts have been identified for 69 of the 73 sources in the two
Hercules fields, and redshifts obtained for 49 of them.  Photometric
redshifts are computed from the $griK$ data for the remaining 21
sources.  This complete sample is compared with the radio luminosity
functions (RLFs) of Dunlop and Peacock~(1990)\cite{Dunlop90}.  The RLF
models successfully trace the evolution of the radio sources with
redshift, but there is some disagreement between the
luminosity-dependence of the models and the data.  The observed RLF
for the lower luminosity population ($\lg P < 26$) shows evidence for
a cut-off at lower redshifts ($z\sim 0.5$--1.5) than for the more
powerful objects.
\end{abstract}

\section*{Introduction}

The purpose of the Leiden--Berkeley Deep Survey (hereafter ``the
LBDS'') was to gain a better understanding of the nature of faint
radio galaxies and quasars, and to determine their cosmological
evolution.  Several high latitude fields in the selected areas SA28,
SA57, SA68 and an area in Hercules had been selected for the purpose
of faint galaxy and quasar photometry, and a collection of good
multi-color prime focus photographic plates had been acquired.  Nine
of these fields were then surveyed with the Westerbork Synthesis Radio
Telescope at 21 cm (1.412 GHz), reaching a 5-$\sigma$ limiting flux
density of 1~mJy \cite{Windhorst84a}.

Following this selection of the radio sample, 171 of the radio sources
(53\%) were identified on the photographic plates, whilst for the
Hercules fields there were 47 out of 73 sources identified
\cite{Windhorst84b,Kron85}.  Presented here are the results of an
extensive optical/infrared investigation of the two Hercules fields,
with the aim of completing the identification and redshift content of
this sub-sample\cite{Waddington98}.  A cosmology of ${\rm
H_0}=50$~\kmsmpc, $\Omega_0=1$ and $\Lambda=0$ is assumed throughout.

\section*{The data}

The Hercules field was observed on the 200~inch Hale telescope at
Palomar Observatory between 1984 and 1988.  Multiple observations were
made through Gunn $g$, $r$ and $i$ filters over the six runs.  After
processing and stacking of the multiple-epoch images, optical
counterparts for 22 of the sources were found, leaving only four
sources unidentified to $r\simeq 26$.  Near-infrared observations have
been made of the entire subsample at $K$, yielding 60/73 detections
down to $K\simeq 19$--21.  Half of the sources have been observed in
$H$ and approximately one-third in $J$.  Observations of the brighter
sources were made by Thuan et~al.~(1984)\cite{Thuan84} and by
Neugebauer et~al.\ and Katgert et~al.\ (priv.\ comm.).  $K$-band
observations of the sample were completed by the present authors at
UKIRT.

Figure~1 presents the optical and infrared magnitude distributions.
For those sources without CCD observations, photographic magnitudes
from Kron et~al.~(1985)\cite{Kron85} have been transformed to the Gunn
system\cite{Windhorst91}.  It is seen that the distribution turns over
at $r \sim 22$, a consequence of evolution in the redshift and/or
luminosity distributions of the radio sources.  The the $r$-band
magnitude distribution is essentially unchanged from this milli-Jansky
survey down to micro-Jansky surveys, a thousand times fainter in radio
flux\cite{Windhorst98}.

\begin{figure}
\centerline{%
\epsfig{file=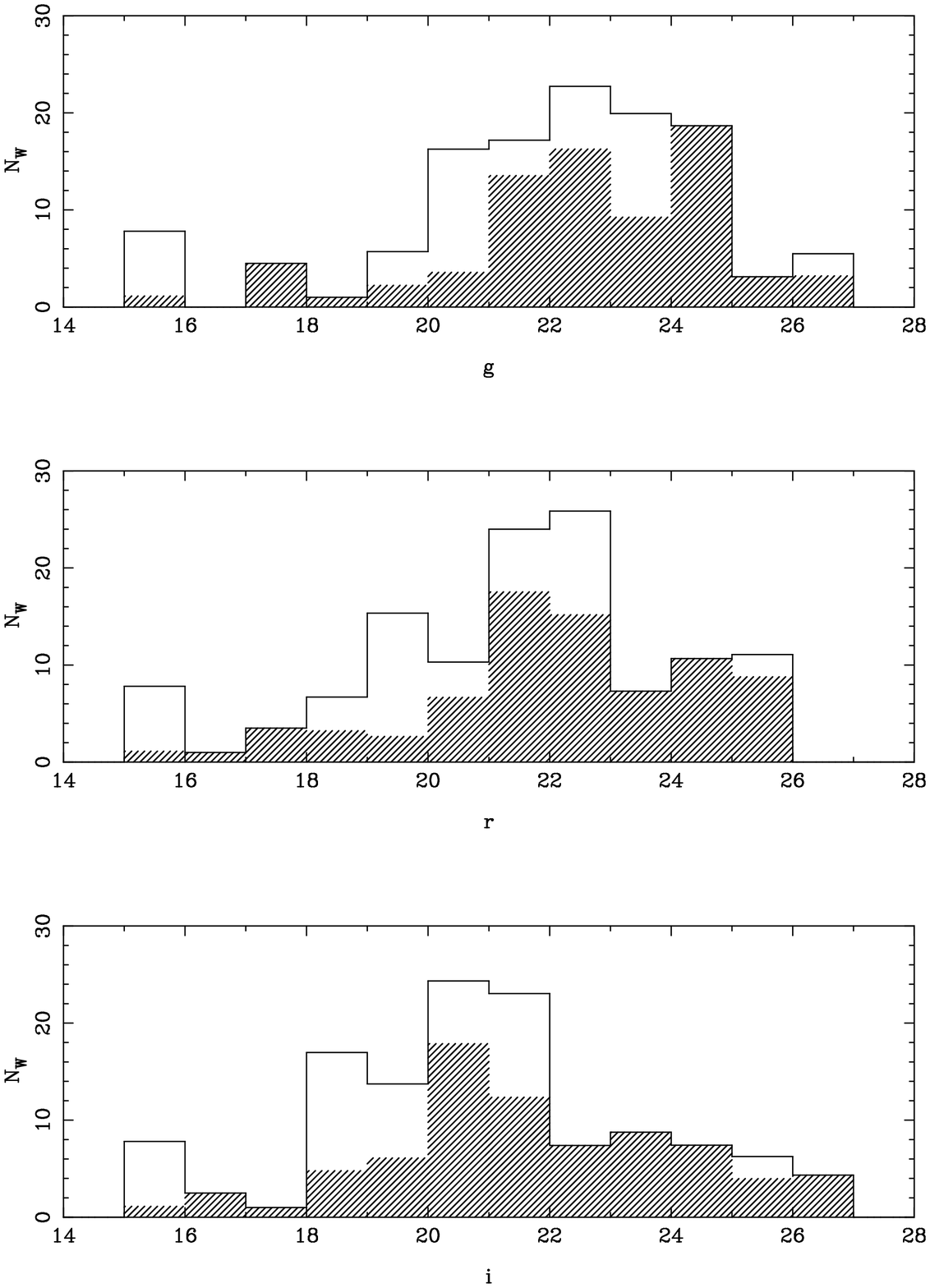,width=7cm,height=8cm} \ \ \ \ \  
\epsfig{file=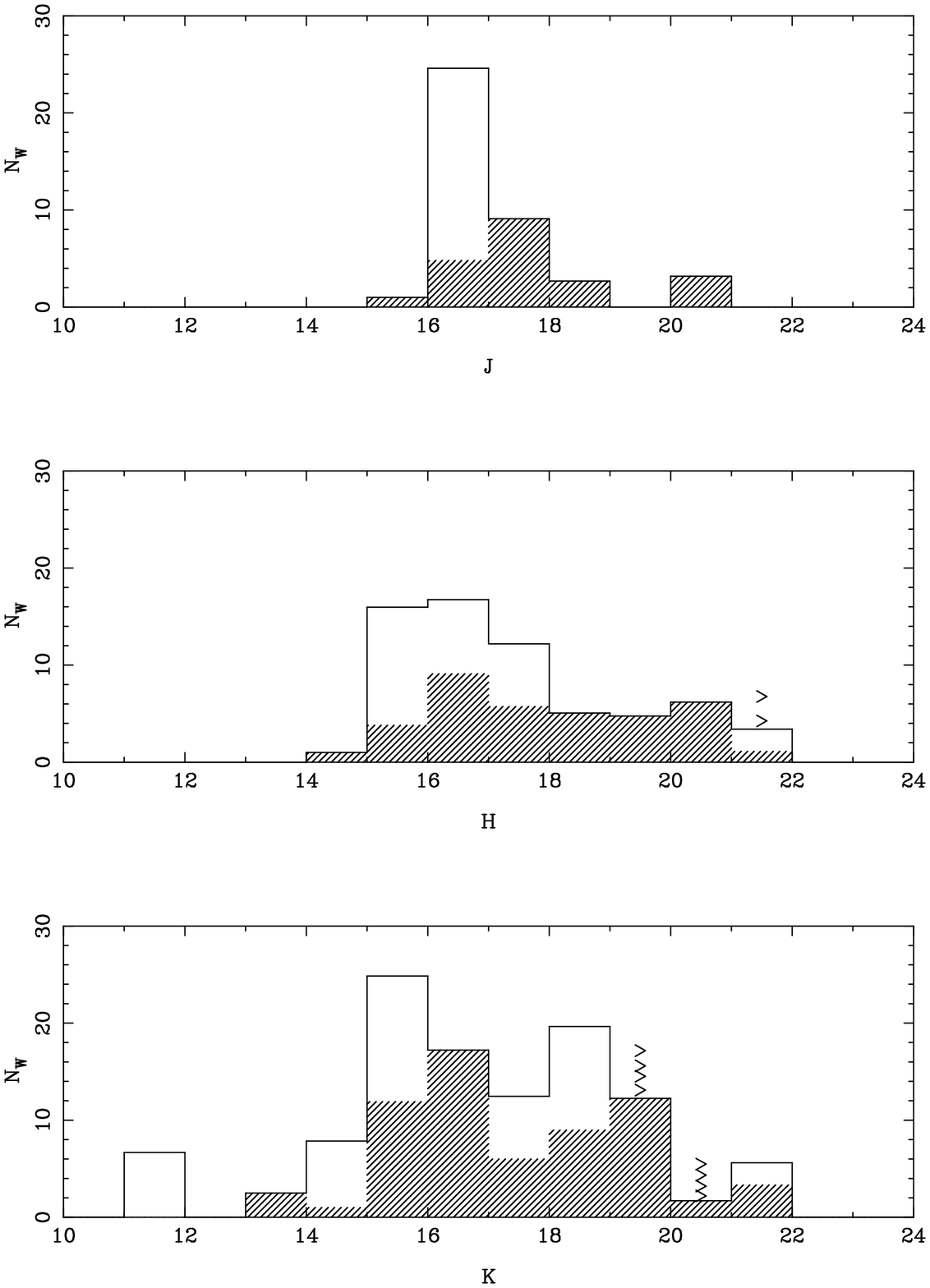,width=7cm,height=8cm}}
\vspace{10pt}
\caption{Magnitude distributions for the LBDS Hercules sample.  Shaded
histograms show the sources with $S_{1.4}\ge 2$~mJy.  Arrows denote
3-$\sigma$ upper limits at $H$ and $K$.}
\label{figureone}
\end{figure}

Prior to the start of the current work, only 16 of the 73 sources in
the LBDS Hercules fields had redshifts published in the literature.
Another 16 sources had unpublished redshifts.  The author and
collaborators have successfully observed a further 17 sources during
the past few years, using both the 4.2~m William Herschel
Telescope\cite{Waddington98} and the 10~m W. M. Keck
Telescope\cite{Dunlop96,Spinrad97,Dey97}.  This brings the total
number of redshifts to 49 out of 73 sources (67\%).

Photometric redshifts were calculated for the remaining one-third of
the sample.  Using the spectral population synthesis models of Jimenez
et~al.~(1998)\cite{Jimenez98}, synthetic $griJHK$ magnitudes were
computed and fitted to the observed magnitudes, giving the
most-probable redshift and a measure of its uncertainty.  Comparison
of the estimated and the true redshifts for those sources with
spectroscopic observations, showed that the average difference was
$\sim 0.1$ in $z$.

\section*{The 1.4~GHz radio luminosity function and the redshift cut-off}

Dunlop and Peacock~(1990)\cite{Dunlop90} used a sample of radio
sources brighter than 0.1~Jy at 2.7~GHz to investigate the radio
luminosity function.  They concluded that the comoving density of both
flat- and steep-spectrum sources suffers a cut-off at redshifts
$z\simeq 2$--4.  This conclusion was drawn from the behavior of both
free-form and simple parametric models (PLE/LDE), and the
model-independent, banded $V/V_{\rm max}$ test.  However, the results
were crucially dependent upon the accuracy of their redshift estimates
in the Parkes Selected Regions (PSR).

\begin{figure}
\centerline{%
\epsfig{file=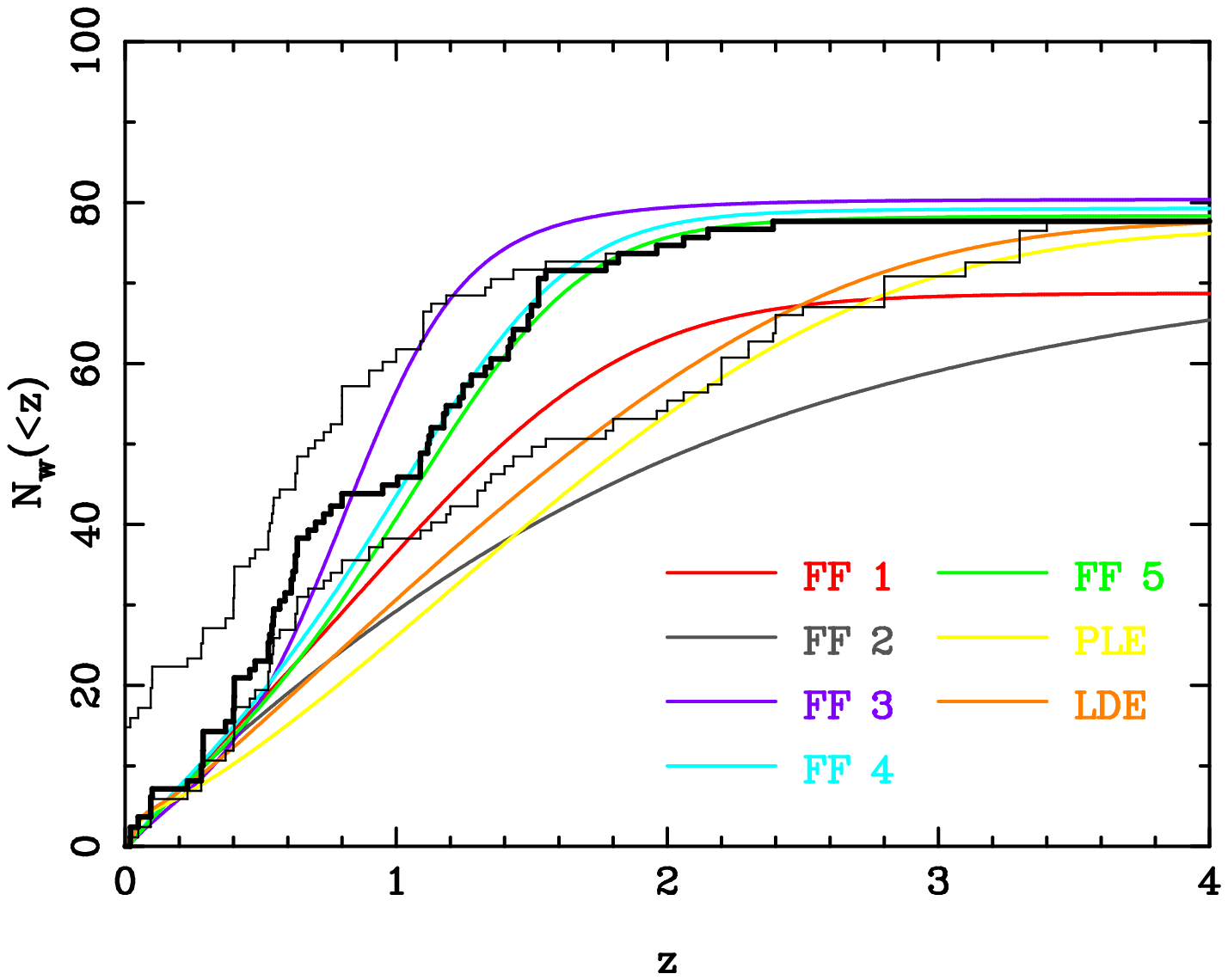,width=8cm,%
height=8cm} \ \ \ \ \  
\epsfig{file=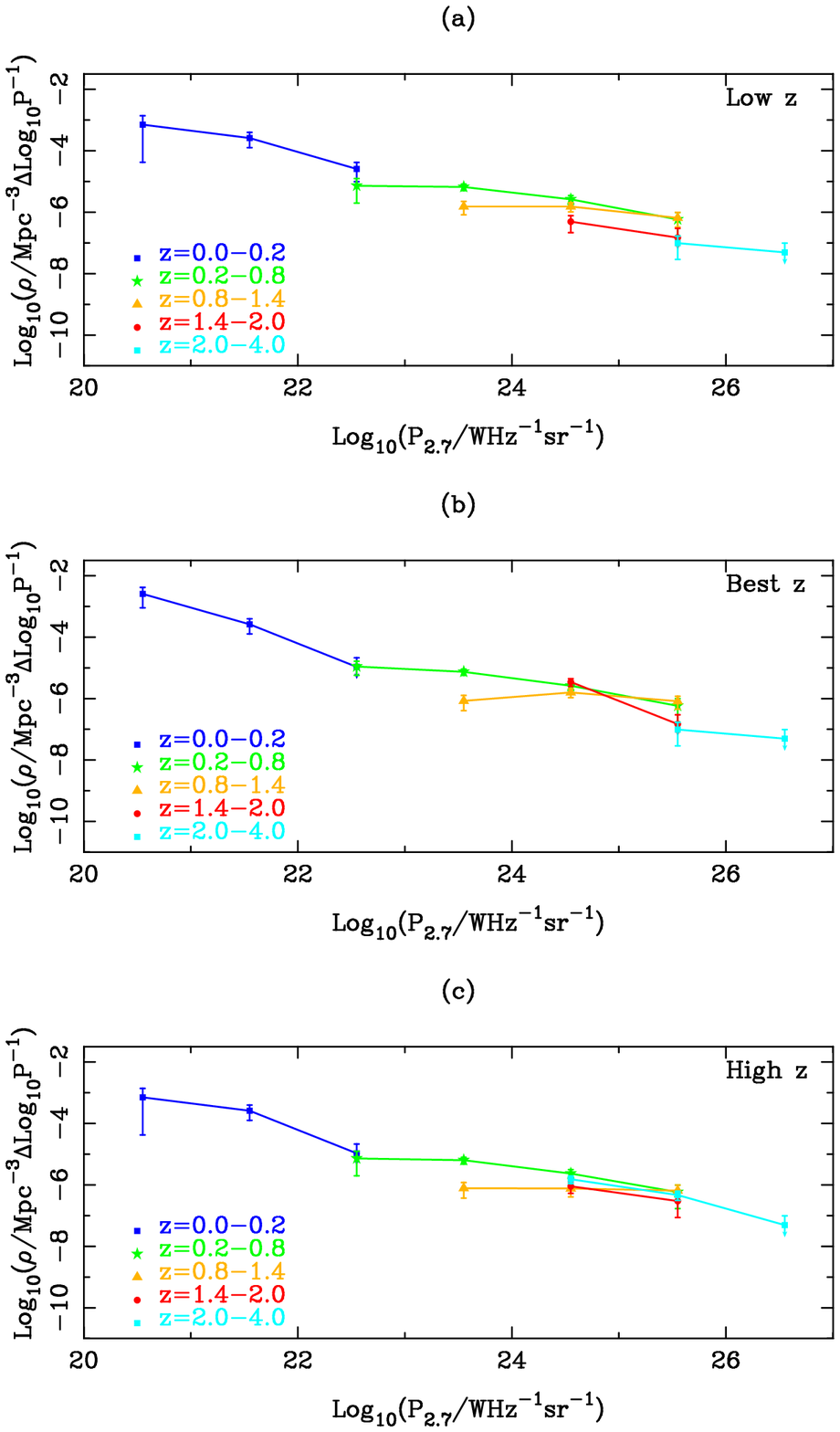,width=6cm,%
height=8cm}}
\vspace{10pt}
\caption{{\bf [Left]} The cumulative redshift distribution of all
sources in the 2-mJy Hercules sample.  The bold histogram is computed
from the best-fit photometric redshift distribution, the lighter
histograms correspond to the lower and upper limits to the photometric
redshifts.  Lines are the model RLFs of [1].  {\bf [Right]} The
observed radio luminosity function for the 2-mJy Hercules sample, for
each of the three photometric redshift distributions.}
\label{figuretwo}
\end{figure}

With a flux limit $\sim 100\times$ fainter than the PSR, the LBDS is
well-suited to test the reliability of those RLF models and the
redshift cut-off, via its potential to detect powerful radio galaxies
at very high redshifts.  In figure~2 [left] the cumulative redshift
distribution of the LBDS Hercules sample (only sources with
$S_{1.4}\ge 2$~mJy) is compared with the predictions of
\cite{Dunlop90}.  It is seen that two of the free-form models (FF-4
and FF-5) provide a reasonable fit to the data over all redshifts.
The ``bump'' in the best-fit histogram at $0.4\la z \la 1$ is due to
two spikes in the redshift distribution, that may be the result of
possible large-scale structures (sheets) along the line of sight.

The observed 1.4~GHz luminosity function presented in figure~2 [right]
was also compared with the models.  It was found that the two models
which fit the cumulative counts (FF-4 and FF-5) do not predict the
observed {\it luminosity\/} dependence of the data nearly as well as
the overall redshift dependence.  The observed RLF shows some
indication that it turns over at $z\simeq 0.5$--1.5, and that the
redshift of this cut-off is a function of the radio luminosity.
However, the small number of sources makes it difficult to separate
the redshift and luminosity dependence of the RLF sufficiently to be
certain of this trend.

The full results of this project are presented in \cite{Waddington98},
and in forthcoming papers by the author and collaborators.

\noindent
{\bf Acknowledgments:} Many people have contributed data and knowledge
to this project.  In particular, I thank James Dunlop, Rogier
Windhorst and John Peacock for their assistance.  The financial
support of the PPARC is acknowledged.

\end{document}